\documentclass{rspublic}
\usepackage{ulem}
\usepackage{graphicx}  %
\usepackage{amsmath}
\usepackage{amsfonts}
\usepackage{bm}  %

\newcommand{\bea}{\begin{eqnarray}}
\newcommand{\eea}{\end{eqnarray}}
\newcommand{\beq}{\begin{equation}}
\newcommand{\eeq}{\end{equation}}
\newcommand{\bqa}{\begin{eqnarray}}
\newcommand{\eqa}{\end{eqnarray}}

\newcommand{\Mhi}{M_{high}}
\newcommand{\Mlo}{M_{low}}
\def\mqo2{{\!\!\!}}

\begin{document}

\title{Efimov physics from a renormalization group perspective}
\author[H.-W. Hammer \& L.~Platter]{Hans-Werner Hammer and Lucas Platter}
\affiliation{Helmholtz-Institut f\"ur Strahlen- und Kernphysik
        (Theorie) and Bethe Center for Theoretical Physics,
        Universit\"at Bonn, 53115 Bonn, Germany\\and\\
Institute for Nuclear Theory, University of Washington, Seattle, WA\ 98195, USA}

\label{firstpage}
\maketitle
\begin{abstract}{Universality, Few-Body Physics}
We discuss the physics of the Efimov effect from
a renormalization group viewpoint using the concept of limit cycles. 
Furthermore, we discuss recent experiments providing evidence for the Efimov 
effect in ultracold gases and its relevance for nuclear systems.
\end{abstract}

\section{Introduction}
\label{sec:introduction}
The {\it Efimov effect} was discovered by Vitaly Efimov
forty years ago (Efimov 1970)
but its significance for low-energy universality
was only realized in the last decade.
It describes the appearance of a discrete scaling
symmetry in the non-relativistic three-body system with a large
two-body scattering length $a$. Efimov showed that
there are infinitely many three-body bound states with an accumulation
point at zero energy in the {\it unitary limit} of infinite 
scattering length
($a\rightarrow\pm\infty$). The ratio of the binding energies of two
successive bound states approaches a universal number $\approx 515$ 
as one approaches the threshold. 

The Efimov effect also has important consequences for finite
scattering length. In particular, it provides the starting point for a
systematic effective field theory description of few-body observables
when the two-body scattering length $a$ is much larger than the range
$\ell$ of the underlying interaction. The effective field theory
approach allows then to include the corrections due to the finite
range of the interaction perturbatively. A more detailed discussion of
these aspects can be found in the reviews by 
Braaten \& Hammer (2006) and Platter (2009).

The Efimov effect has recently received a significant amount of
attention since modern experimental techniques have allowed to
demonstrate its existence in systems that have a large scattering
length. In particular, the ability to tune the scattering length in
experiments with ultracold atoms has allowed to demonstrate the
discrete scaling symmetry by measuring atom loss rates in ultracold
gases (Ferlaino \& Grimm 2010). 
The Efimov effect plays also an important role in nuclear
physics (Hammer \& Platter 2010). 
The nuclei $^3$H and $^3$He can be considered ground
states of an Efimov spectrum with only one state. Some halo nuclei,
which consist of a tightly bound core and a halo of weakly bound
nucleons, are also candidates for Efimov states.

The renormalization group (RG) provides a powerful method 
to understand the physics associated with
the Efimov effect. This tool has improved our understanding of
critical phenomena in condensed matter physics and of the asymptotic
behavior of field theories in particle physics (Wilson 1983). In this
review, we discuss recent progress in the theoretical description and
experimental observation of the Efimov effect. We emphasize the
importance of RG concepts as a tool for understanding Efimov
physics. In sec.~\ref{sec:renorm-group}, we briefly introduce the
concept of the renormalization group. We then discuss the connection
between renormalization group {\it fixed points} and {\it limit
  cycles} to continuous and discrete scaling symmetries. These ideas
are illustrated using the inverse square potential as an example.  In
sec.~\ref{sec:efimov-effect}, we demonstrate that the Efimov effect is
a result of an infrared limit cycle and discuss some of the observable
features associated with the term {\it Efimov physics}.
The consequences of Efimov physics in the four-body system
are discussed in sec.~\ref{sec:four-body-legacy}. In 
sec.~\ref{sec:physical-systems}, we give a broad overview of Efimov physics 
in systems from atomic to nuclear physics.
We end with conclusions and an outlook.

\section{The Renormalization Group}
\label{sec:renorm-group}

The renormalization group (RG) is one of the most important
developments in modern physics (Wilson 1983).
The RG has been particularly important
for the understanding of {\it critical phenomena}. This term is used
to describe universal behavior in second order phase transitions.
Near a {\it critical point} certain thermodynamic observables show
power law behavior. This power law behavior at the critical point
implies scale invariance of the system. For example, the correlation
length $\xi$ diverges as $T \to T_c$:
\begin{equation}
\label{eq:corr}
\xi(T) \to {\rm const.}\, |T-T_c|^{-\nu}\,,
\end{equation}
where $T_c$ is the critical temperature and $\nu$ is a {\it critical
  exponent}. As a consequence, the system looks the same at all
distance scales.  The critical exponents are the same for systems in
the same {\it universality class}. 
An example of two such systems in the
same universality class are liquid-gas systems and a ferromagnet with
one easy axis of magnetization. Even though their dynamics at
short distances is very different, 
they share the same critical exponents.  From the
perspective of the renormalization group, this scaling behavior is
characterized by a fixed point. Such fixed points also play an
important role in quantum field theories in particle physics.
Asymptotically free theories like Quantum Chromodynamics, e.g., have
an ultraviolet fixed point that corresponds to zero coupling
(Gross \& Wilczek 1973; Politzer 1973).

In order to understand the concept of renormalization group fixed
points, it is convenient to expand the Hamiltonian $\mathcal{H}$
governing the system in a basis of operators $\mathcal{O}_n$
\begin{eqnarray}
  \label{eq:4}
  \mathcal{H}=\sum_n g_n \mathcal{O}_n~\,.
\end{eqnarray}
We refer to the basis coefficients $g_n$ as the coupling constants
of the Hamiltonian. A particular Hamiltonian is represented by a point
${\bf g}=(g_1, g_2,\ldots)$ in the infinite dimensional vector space.
If we now introduce an ultraviolet cutoff $\Lambda$, an RG
transformation is defined as a transformation that eliminates
short-distance degrees of freedom (lowers the cutoff $\Lambda$) while
the long-distance observables remain unchanged. We treat the cutoff $\Lambda$
as a continuous variable and the RG transformation defines thus a flow
in the space of coupling constants. This flow can be expressed through
a differential equation for the coupling ${\bf g}$
\begin{eqnarray}
  \label{eq:5}
  \beta({\bf g})=\Lambda \frac{\hbox{d}}{\hbox{d}\Lambda}{\bf g}~,
\end{eqnarray}
where $\beta({\bf g})$ is the {\it beta function}.
This equation can be used to define the concept of a fixed point. A
fixed point ${\bf g}_*$ in the space of coupling constants fulfills
\begin{eqnarray}
  \label{eq:6}
  \beta({\bf g_*})=0~.
\end{eqnarray}
Wilson discovered that scale invariant behavior at long distances is a
result of a fixed point in the infrared limit $\Lambda\rightarrow 0$
(infrared fixed point). The Hamiltonian $\mathcal{H}_*$ remains then
invariant under a change of the ultraviolet cutoff $\Lambda$. Since
$\Lambda$ defines the length scale of the system, the system is
therefore scale invariant at the fixed point.

Equation \eqref{eq:6} can be linearized in the vicinity of a fixed
point 
\begin{eqnarray}
  \label{eq:7}
  \Lambda\frac{\hbox{d}}{\hbox{d}\Lambda}{\bf g}\approx B({\bf g}-{\bf g}_*)~,
\end{eqnarray}
where $B$ is a linear operator. The eigenvalues of the operator $B$
are the critical exponents. Operators with positive, zero
and negative critical exponents are called relevant, marginal and
irrelevant operators, respectively. Relevant operators become more
important for large $\Lambda$, while irrelevant operators become less
important. Marginal operators are a special case and higher order
corrections are important to decide their fate. The critical exponent
$\nu$ in Eq.~\eqref{eq:corr} is related to the
critical exponent of an appropriate operator in the Hamiltonian.

Another solution to Eq. \eqref{eq:5}, corresponds to a {\it limit
  cycle}. In this case the RG trajectory flows around a closed loop in
the space of coupling constants. A limit cycle can be understood as a
family of Hamiltonians $\mathcal{H}_*(\theta)$ that is closed under
the RG flow and can be parameterized by an angle $\theta$ that runs
from 0 to $2\pi$. All coupling constants in the Hamiltonian
$\mathcal{H}$ will return to their initial value once the cutoff is
changed by a factor $\mathcal{S}_0$ (the {\it
  preferred scaling factor}):
\begin{eqnarray}
  \label{eq:8}
  {\bf g}(\Lambda)={\bf g}_*(\theta_0 +2\pi \ln(\Lambda/\Lambda_0)
/\ln(\mathcal{S}_0))~.
\end{eqnarray}
A necessary condition for a limit cycle is invariance under discrete scale
transformations: $x\to (\mathcal{S}_0)^n x$, where $n$ is an integer. This
discrete scaling symmetry is reflected in log-periodic behavior of
physical observables.
With a single coupling constant, a limit cycle can only occur if the
coupling has discontinuities. If this is the case, the RG equation 
has two complex conjugate fixed point solutions,
see e.g. (Moroz \& Schmidt 2010).
The preferred scaling factor $\mathcal{S}_0$ is then 
determined by the imaginary part of the fixed point coupling.

The possibility of RG limit cycles was first discussed by Wilson in a
work that applied the concepts of the renormalization group
to the strong interactions of elementary particle physics (Wilson 1971). 
The Efimov effect can be understood
in terms of a renormalization group limit cycle with
discrete scaling factor $\mathcal{S}_0 \simeq 22.7$ (Albeverio {\it et
  al.} 1981).
In the limit $a \to \pm \infty$, there is an accumulation of
three-body bound states near threshold with binding energies differing by
multiplicative factors of $(\mathcal{S}_0)^2 \simeq 515.03$ (Efimov 1970).
In the effective field theory formulation of Bedaque {\it et al.} (1999), the
limit cycle becomes explicit in the running of the leading-order three-body 
contact interaction required for renormalization.
A renormalization group analysis of the three-body problem was carried out
by Barford \& Birse (2005).

We will use the quantum mechanical $1/r^2$ potential as an example 
that exhibits both fixed point and limit cycle solutions. 
Our discussion follows Kaplan {\it et al.} (2009) and Hammer \& Swingle (2006).
Consider the three-dimensional Schr\"odinger equation with the potential $V(r)$
\begin{equation}
  \label{eq:inv-square-pot}
  V(r)=\frac{\alpha}{r^2}~,
\end{equation}
where we have set $\hbar=m=1$ for convenience, such that $\alpha$ is
dimensionless.
This potential is scale invariant at the classical level.
Rescaling all distances by a factor $\lambda$ 
and rescaling time by a factor of $\lambda^2$ leaves the Hamiltonian 
invariant up to a constant prefactor $1/\lambda^2$.

It is straightforward
to show that the $S$-wave solution to the Schr\"odinger equation 
for $\alpha_*<\alpha<\alpha_*+1$ and zero energy is given by
\begin{equation}
  \label{eq:invsq-solution}
  \psi(r)=c_{-}r^{\nu_{-}}+c_{+}r^{\nu_{+}}~,\quad
  \nu_{\pm}=-\frac{1}{2}\pm\sqrt{\alpha-\alpha_*}~,
\end{equation}
where \
\begin{equation}
  \label{eq:alphastar}
  \alpha_*\equiv-\frac{1}{4}~.
\end{equation}
If either $c_-=0$ or $c_+=0$, this solution is scale invariant. 
We can identify the solution $c_-=0$ with an 
infrared fixed point and the solution $c_+=0$ with an ultraviolet 
fixed point. Formally, one can pick the desired solution by 
adding a counterterm potential with coupling
constant $g$ to the Hamiltonian.
This is equivalent to putting a boundary condition on the 
wave function at short distances.
The coupling constant $g$ satisfies an RG equation with a 
$\beta$-function that takes the form (Kaplan {\it et al.} 2009)
\begin{equation}
\beta(g;\alpha)=(\alpha-\alpha_*)-(g-g_*)^2~.
\label{eq:betaf}
\end{equation}
For $\alpha > \alpha_*$, the fixed points of $g$ are
\begin{equation}
g_{\pm}=g_* \pm \sqrt{\alpha-\alpha_*}\,,
\label{eq:fixp}
\end{equation}
where $g_-$ and $g_+$ correspond to the infrared and ultraviolet
fixed points for the solutions
with $c_-=0$ and $c_+=0$, respectively.

As $\alpha$ is decreased and approaches $\alpha_*$ from above, the two
fixed points approach each other and merge at $g_\pm=g_*$. 
For $\alpha < \alpha_*$, the fixed point equation $\beta(g;\alpha)=0$
has two complex conjugate solutions 
\begin{equation}
g_{\pm}=g_* \pm i \sqrt{\alpha_*-\alpha}\,,
\label{eq:fixpc}
\end{equation}
which correspond to a limit cycle with
preferred scaling factor
\begin{equation}
\mathcal{S}_0=\exp(\pi/\sqrt{\alpha_*-\alpha})\,.
\label{eq:lam0}
\end{equation}
The continuous scaling symmetry of the system is now broken to a 
discrete scaling symmetry by quantum fluctuations at small distances.
This symmetry breaking
can be interpreted as a quantum mechanical anomaly (Ananos {\it
  et al.} 2003).
The discrete scaling symmetry is manifest in observables such 
as in the geometric bound state spectrum 
\begin{equation}
B^{(n)} = B_0 (\mathcal{S}_0)^{-2n}\,,
\end{equation}
as well as the log-periodic dependence of the $S$-wave scattering
phase shift
\begin{equation}
\delta(k)= \delta_0 - \frac{\pi\ln k}{\ln\mathcal{S}_0}\,,
\end{equation}
and the corresponding cross section on the momentum $k$.

\section{Efimov Effect}
\label{sec:efimov-effect}
\begin{figure}[t]
\centerline{\includegraphics*[width=8cm,angle=0]{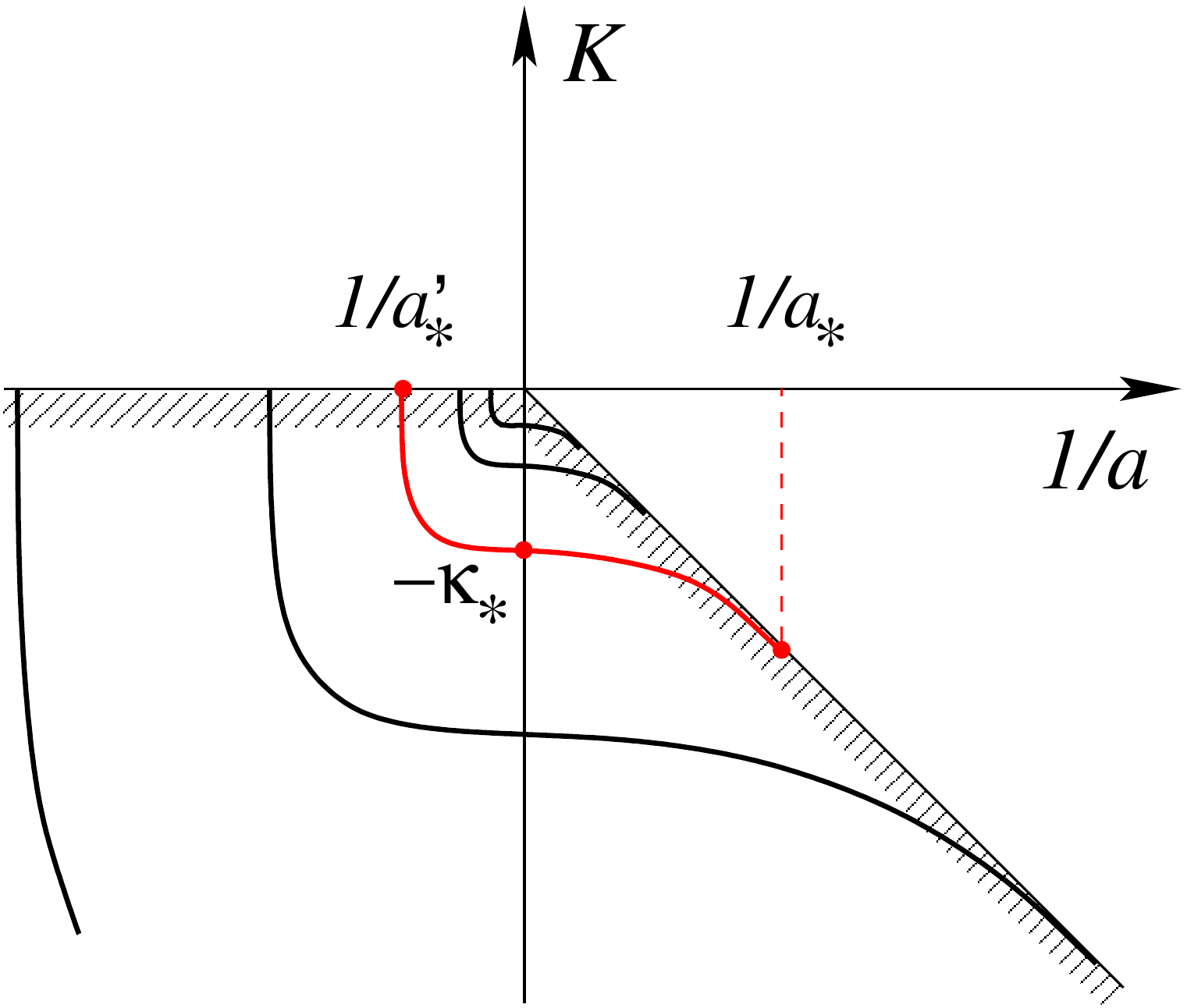}}
\caption
{The Efimov plot for the three-body problem. We show
$K\equiv {\rm sgn}(E)(m|E|)^{1/2}/\hbar$ versus the inverse scattering 
length $1/a$. The solid lines denote some of 
the infinitely many branches of Efimov states while
the cross-hatching indicates the threshold for scattering states.
Either one of the quantities $a_*$, $a'_*$, and $\kappa_*$ can be
used to specify the branch labelled with $n=n_*$. All other states
follow from the discrete scaling symmetry.
}
\label{fig:efimovplot}
\end{figure}

In 1970, Efimov analyzed the three-nucleon system interacting through
zero-range interactions. This system had been considered before, but he
was the first one to realize that one should not focus on the
pathologies at large energies but on the universal physics at
low energies, $E \ll \hbar^2/m\ell^2$.  In this limit, where
zero-range forces are adequate, he found some surprising results
(Efimov 1970).  He pointed out that when $|a|$ is sufficiently
large compared to the range $\ell$ of the potential, there is a
sequence of three-body bound states whose binding energies are spaced
roughly geometrically in the interval between $\hbar^2/m \ell^2$ and
$\hbar^2/m a^2$.  As $|a|$ is increased, new bound states appear in
the spectrum at critical values of $a$ that differ by multiplicative
factors of $e^{\pi/s_0}$, 
where $s_0$ depends on the statistics and
the mass ratios of the particles.  In the case of spin-doublet
neutron-deuteron scattering and for three identical bosons, $s_0$ is
the solution to the transcendental equation
\begin{eqnarray}
s_0 \cosh {\pi s_0 \over 2} = {8 \over \sqrt{3}} \sinh {\pi s_0 \over 6} \,.
\label{s0}
\end{eqnarray}
Its numerical value is $s_0\approx 1.00624$ and the preferred scaling factor
is
\begin{equation}
\mathcal{S}_0 =e^{\pi/s_0}\approx
22.7\,.
\end{equation}  
As $|a|/\ell \to \infty$, the asymptotic number of three-body bound
states is
\begin{eqnarray}
N \longrightarrow {1 \over \ln\mathcal{S}_0} \ln {|a| \over \ell} \,.
\label{N-Efimov}
\end{eqnarray}
Equation \eqref{N-Efimov} expresses the fact that in any system with a
finite range, the number of Efimov states is bounded from above due to the
finite range of the interaction. Since Efimov's approach requires the
scattering length $a$ to be significantly
larger than the range of the interaction $\ell$,
Efimov states must have binding energies smaller than $\hbar/m\ell^2$. The
three-nucleon systems $^3$H or $^3$He contain therefore only one Efimov state
which is also the ground state.
In the limit $a \to \pm \infty$, there are infinitely 
many three-body bound states with an accumulation point at the 
three-body scattering threshold with a geometric spectrum:
\begin{eqnarray}
B^{(n)}_t = \mathcal{S}_0^{-2(n-n_*)} \hbar^2 \kappa^2_* /m,
\label{kappa-star}
\end{eqnarray}
where $m$ is the mass of the particles and $\kappa_*$ is the binding
wavenumber of the branch of Efimov states labeled by $n_*$
(See Fig.~\ref{fig:efimovplot}).  The
geometric spectrum in (\ref{kappa-star}) is the signature of a
discrete scaling symmetry with scaling factor 
$\mathcal{S}_0 \approx 22.7$.  
It is independent of the mass or structure of the identical
particles and independent of the form of their short-range
interactions.  The Efimov effect can also occur in other three-body
systems if at least two of the three pairs have a large S-wave
scattering length but the numerical value of the asymptotic ratio may
differ from the value $22.7^2\approx 515$.

A formal proof of the Efimov effect was subsequently given by Amado
and Noble (Amado \& Noble 1971, 1972).  The Thomas and Efimov
effects are closely related. The deepest three-body bound states found by
Thomas' variational calculation can be identified with the deepest
Efimov states (Adhikari {\it et al.} 1988).

The universal properties in the three-body system with large scattering
length are not restricted to the Efimov effect. The dependence of
three-body observables on the scattering length is
characterized by scaling behavior modulo coefficients that are
log-periodic functions of $a$ (Efimov 1971, 1979).
This behavior is characteristic of a system with a discrete scaling
symmetry.  We will refer to universal aspects associated with a
discrete scaling symmetry as Efimov physics.

In 1981, Efimov proposed a new approach to the low-energy few-nucleon
problem in nuclear physics that, in modern language, was based on
perturbation theory around the unitary limit (Efimov 1981).
Remarkably, this program works reasonably well in the three-nucleon system
at momenta small compared to $M_\pi$. It has been used as the basis
of an effective field theory for nuclear physics at very low energies
(Bedaque \& van Kolck 2002; Epelbaum {\it et al.} 2009).

Efimov formulated the three-body problem using hyperspherical
coordinates that are particularly well-suited for the analysis of
few-body problems in coordinate space. In this approach, the 
6 independent coordinates of the problem are given by the 
hyperradius and 5 angles. The hyperradius $R$ is defined as
\begin{equation}
 R^2=\frac{1}{3}({\bf r}_{12}^2 + {\bf r}_{23}^2 + {\bf r}_{31}^2)~,
\end{equation}
where ${\bf r}_{ij}={\bf r}_{i}-{\bf r}_{j}$ is the separation between
atoms $i$ and $j$. It is only small if all three
particles are close together.
Efimov showed that in these coordinates the three-body problem with a
large scattering length reduces to a simple Schr\"odinger-like
equation with a $1/R^2$ potential. At short distances, the problem 
simplifies and the radial equation takes the form
\begin{equation}
  \frac{\hbar^2}{2 m}\left(-\frac{\partial^2}{\partial R^2} - \frac{s_0^2 +
  \frac{1}{4}}{R^2}\right)f_0(R)=Ef_0(R)~,\qquad R\ll a\,.
\label{eq:hyperradeqLO}
\end{equation}
As discussed in the previous section, the $1/R^2$ potential requires
a boundary condition on the wave function at short distances 
if the potential is sufficiently attractive.
This boundary condition, which alternatively
can be expressed through a short-range three-body force
(Bedaque {\it et al.} 1999), can be fixed from any three-body observable.

In practical applications to the three-nucleon system,
one uses either the spin-doublet neutron-deuteron scattering
length or the triton binding energy as three-body input. 
If the deuteron binding
energy and the spin-singlet scattering length are used as two-body input
and if the boundary condition is fixed using the spin-doublet 
neutron-deuteron scattering length, the triton binding energy is 
predicted with an accuracy of 6\%
(Bedaque {\it et al.} 2000). The accuracy of 
the predictions can be further improved by taking
into account the effective range as a first-order perturbation
(Efimov 1991). Thus, the triton can be identified as an Efimov
state associated with the deuteron being a $pn$ bound state with large
scattering length (Efimov 1981).

The effective coupling of the potential in Eq.~\eqref{eq:hyperradeqLO}
$\alpha=-(s_0^2+1/4)$ is smaller than $\alpha_*=-1/4$. 
This implies that the renormalization group flow of
the counterterm introduced for regularization of the problem will
display limit cycle behaviour. This limit cycle has important
consequences for observables. For example in the unitary limit the
binding energies scale as given in Eq.~\eqref{kappa-star}.  For finite
scattering length, observables such as binding energy and cross
sections still scale with integer powers of
$\mathcal{S}_0=\exp(\pi/s_0)$ under this symmetry.  For example, the
binding energy of an Efimov trimer which is a function of $a$ and
$\kappa_*$ scales
\begin{equation}
  \label{eq:1}
  B_3^{(n)}(\mathcal{S}_0^m a,\kappa_*)=\mathcal{S}_0^{-2m}B_3^{(n-m)}(a,\kappa_*)~.
\end{equation}
This implies for positive scattering length
\begin{equation}
  \label{eq:2}
  B_3^{(n)}(a,\kappa_*)=F_n (2 s_0\ln(a \kappa_*))\frac{\hbar^2\kappa_*^2}{m}~.
\end{equation}
The function $F_n$ parameterizes the scattering length dependence of
all Efimov trimers exactly in the limit of vanishing range. The
function $F_n$ satisfies
\begin{equation}
  \label{eq:3}
  F_n(x+2m\pi)=\mathcal{S}_0^{-2m} F_{n-m}(x)~.
\end{equation}
The scattering length dependence of the bound state spectrum is shown
in Fig. \ref{fig:efimovplot}. We plot the quantity $K\equiv {\rm
  sgn}(E)(m|E|)^{1/2}/\hbar$ against the inverse scattering length.
For bound states $K$ corresponds to the binding momentum.  The solid lines
denote the Efimov trimers while the scattering threshold  
is indicated by the hatched area. For $a<0$ the relevant threshold is the 
three-particle threshold. For $a>0$, the atom-dimer threshold is lower 
in energy. Only a
few of the infinitely many Efimov branches are shown. A given physical
system has a fixed scattering length value and the corresponding states
lie on a vertical line.  Changing the parameter $\kappa_*$ by a factor
$\mathcal{S}_0$ corresponds to multiplying each branch of trimers with
this factor without changing their shapes.  One important result is
that three-bound states exist for positive and negative scattering
length. This is remarkable for the latter case since the two-body
subsystem is unbound for $a<0$. At a negative scattering length 
denoted by $a'_*$, 
a bound state with a given $\kappa_*$ has zero
binding energy. As the inverse 
scattering length is increased, the trimer binding
energy gets larger until it crosses the atom-dimer threshold at the
positive scattering length $a_*$. The quantities $a_*$, $a'_*$, and
$\kappa_*$ are related to each other and can
be used to quantify a branch of Efimov states. The other branches 
then follow from the discrete scaling symmetry.

\section{The Four-Body Legacy and Beyond}
\label{sec:four-body-legacy}
We discussed in the previous section that one three-body parameter is
required for a consistent description of the three-body system with
zero-range interactions. It is therefore natural to ask how many
parameters are needed for calculations in the $N$-body system. A first
step towards answering this question was performed in the work by
Platter {\it et al.} (2004). 
The authors of this work showed that the two-body scattering
length and one three-body parameter are sufficient to make predictions
for four-body observables. Results of a more detailed analysis (Hammer
\& Platter 2006) also lead to the conclusion that every trimer state is tied to
two universal tetramer states with binding energies related to the
binding energy of the next shallower trimer. In the unitary limit
$1/a=0$, the relation between the tetramer and trimer binding
energies was found as:
\begin{equation}
  \label{eq:4body1}
  B_4^ {(0)}\approx 5\, B_3\quad {\rm and} \quad  B_4^{(1)}\approx 1.01\, B_3~,
\end{equation}
where $B_4^ {(0)}$ denotes the binding energy of the deeper of the two tetramer
states and $B_4^ {(1)}$ the shallower of the two.

A recent calculation by von Stecher {\it et al.} (2009) supports these
findings and extends them to higher numerical accuracy.  For the
relation between universal three- and four-body bound states in the
unitary limit, they found
 \begin{equation}
  \label{eq:4body2}
 B_4^ {(0)}\approx 4.57\, B_3\quad {\rm and} \quad  B_4^{(1)}\approx 1.01\, B_3~,
\end{equation}
which is consistent with the results given in Eq.~(\ref{eq:4body1})
within the numerical accuracy. The universality of the deeper
tetramer was also confirmed in a functional RG analysis by
Schmidt \& Moroz (2010), but the shallower tetramer could not be resolved.

\begin{figure}[tb]
\centerline{\includegraphics*[width=10cm,angle=0]{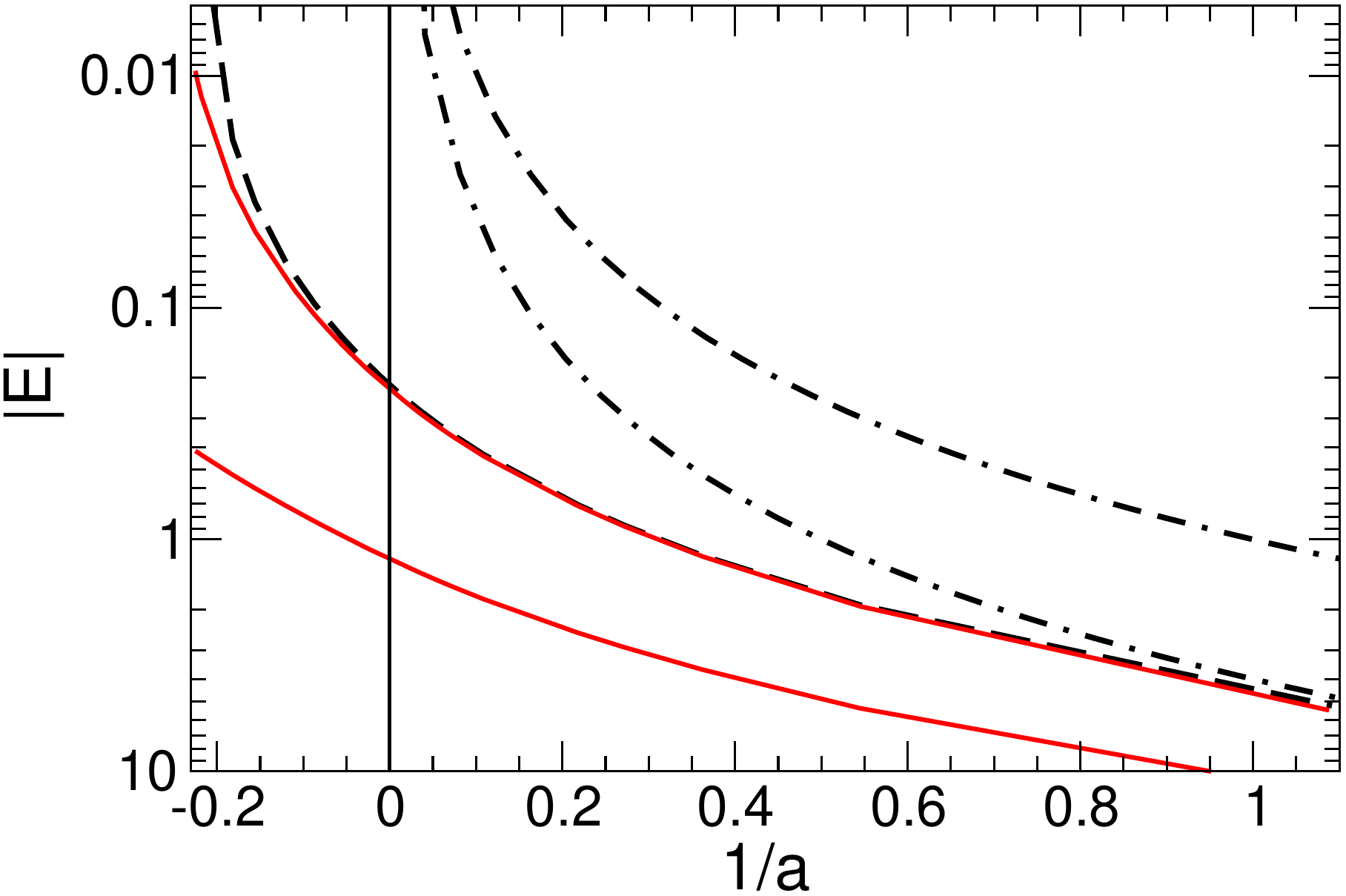}}
\caption{\label{fig:efimov-4body} 
The extended Efimov plot for the four-body problem. We show
$K\equiv {\rm sgn}(E)(m|E|)^{1/2}/\hbar$ versus the inverse scattering 
length $1/a$. Both quantities are given in arbitrary units.
The lower and upper solid lines indicate the four-body ground and excited
states, respectively, while the dashed
line gives the threshold for decay into a ground state
trimer and a particle. The dash-dotted (dash-dash-dotted) lines give 
the thresholds for decay into two dimers (a dimer and two particles).  
}
\end{figure}

The results obtained by the Hammer and Platter (Hammer \& Platter 2006) were
furthermore presented in the form of an extended Efimov plot, shown in
Fig.~\ref{fig:efimov-4body}. Four-body states have to have a binding
energy larger than that of the deepest trimer state. The
corresponding threshold is denoted by the dashed line in
Fig.~\ref{fig:efimov-4body}.  
At positive scattering length, there are also scattering
thresholds for scattering of two dimers and scattering of a dimer and
two particles indicated by the dash-dotted and dash-dash-dotted lines,
respectively.  

An extended version of this four-body Efimov plot was presented
by von Stecher {\it et al.} (2009).  They
calculated more states with higher numerical accuracy and extended the
calculation of the four-body states to the thresholds where they
become unstable.
From these results they extracted the negative values of the 
scattering lengths at which the binding energies of the tetramer 
states become zero and found
\begin{equation}
\label{eq:tetra-scatteringlengths}
  a^*_{4,0}\approx 0.43 a'_*\quad{\rm and}\quad a^*_{4,1}\approx 0.92 a'_*~.
\end{equation}
These numbers uniquely specify the relative position of three- and
four-body recombination resonances. This was the key information for
the subsequent observation of these states in ultracold atoms by
Ferlaino {\it et al.} (2009).

Recently, Deltuva (2010) calculated four-body scattering
observables. He solved the atom-trimer scattering
problem in momentum space and calculated the real and imaginary
parts of the atom-trimer scattering length and effective range:
\begin{equation}
  \label{eq:4bodyscat}
  a_T=(18.5+i\,0.890) \kappa_*^{-1}~\quad{\rm and}
  \quad r_T=(2.63+i\,0.014) \kappa_*^{-1}~.
\end{equation}
He also obtained more precise values for the universal relations
between an Efimov trimer and the associated tetramers:
\begin{equation}
  \label{eq:4body3}
 B_4^ {(0)}\approx 4.6108\, B_3\quad {\rm and} \quad  
 B_4^{(1)}\approx 1.00228\, B_3~.
\end{equation} 
Interestingly, the widths of the excited tetramer states are also
universal. They are 0.3\% of the binding energy for the deeper
state and 0.02\% of the binding energy for the shallower state (Deltuva 2010).

The bound state properties of larger systems of bosons interacting
through short-range interactions were considered by Hanna \& Blume
(2006). Using Monte Carlo methods they showed that universal
correlations between binding energies can also be obtained for those.

Calculations for larger number of particles using a model that
incorporates the universal behavior of the three-body system were
carried our by von Stecher (von Stecher 2009b). His findings indicate
that there is at least one $N$-body state tied to each Efimov
trimer and numerical evidence was also found for a second excited
5-body state.

\section{Physical Systems}
\label{sec:physical-systems}
Efimov's work was originally intended as a description of the
three-nucleon system. However, experiments with ultracold atoms have
proven to be an indispensable testing ground for universal
predictions. Using {\it Feshbach resonances}, the scattering 
length of the atoms can be tuned experimentally by varying an
external magnetic field (see Chin {\it et al.} (2010) for
more details). As a consequence, the discrete scaling symmetry
can be tested by measuring the scattering length dependence of 
observables.

In the following two subsections, we first summarize recent
experimental efforts that have provided evidence for the Efimov effect
in ultracold atoms and then discuss nuclear systems in which 
Efimov physics is expected to play an important role.
\subsection{Atomic Physics}
The first experimental evidence for Efimov physics in ultracold atoms
was presented by Kraemer {\it et al.}\ (2006). This
group used $^{133}$Cs atoms in the lowest hyperfine spin state. They
observed a resonant enhancement of the loss of atoms from three-body
recombination that can be attributed to an Efimov trimer crossing the
three-atom threshold. Such crossings occur when the scattering length
is equal to $(\mathcal{S}_0)^n\,a'_*$ with $n$ an integer 
(cf.~Fig.~\ref{fig:efimovplot}).
The occurrence of recombination resonances 
at negative scattering length was predicted by Esry {\it et al.}\ (1999). 
The universal line shape for the resonance as a
function of the scattering length was first derived by Braaten \&
Hammer (2004). Kraemer {\it et al.}\ also observed a minimum
in the three-body recombination rate that can be interpreted as an
interference effect associated with Efimov physics. 
One of the most exciting developments in the field of Efimov physics
involves universal tetramer states. Ferlaino {\it et al.}\ recently
observed two tetramers in an ultracold gas of $^{133}$Cs atoms
(Ferlaino {\it et al.} 2009) and confirmed the prediction of such states
by Platter {\it  et al.} (2004), Platter \& Hammer (2006) and von 
Stecher {\it et  al.} (2009).

Recent experiments with other bosonic atoms have provided even
stronger evidence of Efimov physics in the three- and four-body
sectors. Zaccanti {\it et al.}\ (2008) measured the three-body 
recombination rate and
the atom-dimer loss rate in a ultracold gas of $^{39}$K atoms. 
They observed two atom-dimer loss resonances and two
minima in the three-body recombination rate at large positive values
of the scattering length.  The positions of the loss features are
consistent with the universal predictions with discrete scaling factor
22.7. They also observed loss features at large negative scattering
lengths. Barontini {\it et al.} (2009) obtained the first evidence 
of the Efimov
effect in a heteronuclear mixture of $^{41}$K and $^{87}$Rb atoms. 
They observed three-atom loss resonances at large
negative scattering lengths in both the K-Rb-Rb and K-K-Rb channels,
for which the discrete scaling factors are 131 and $3.51 \times 10^5$,
respectively. A theoretical analysis of the heteronuclear case was 
presented in (Helfrich {\it et al.} 2010).

Gross {\it et al.} (2009) measured the three-body recombination
rate in an ultracold system of $^7$Li atoms. They
observed a three-atom loss resonance at a large negative scattering length
and a three-body recombination minimum at a large positive scattering
length. The positions of the loss features, which are in the same
universal region on different sides of a Feshbach resonance, are
consistent with the universal predictions with discrete scaling factor
22.7. Pollack {\it et al.} (2009) at Rice University measured 
the three-body recombination in a system of
$^7$Li atoms in a different hyperfine state.
They observed a total of 11 three- and
four-body loss features. The features obey the universal relations on
each side of the Feshbach resonance separately, however, a systematic
error of $\sim 50$ \% is found when features on different sides of the
Feshbach resonance are compared. Recently, the Rice measurement
was repeated by Gross {\it et al.} (2010). They found that recombination 
features were related across the
Feshbach resonance as predicted by the universal relations. Gross
  {\it et al.} also remeasured the magnetic field strength at which the
resonance occurs and found their result to be different from Pollack
{\it et al.} (2009). They claim that this discrepancy might explain
the deviations from universality in the Rice experiment.

Efimov physics has also been observed in the fermionic three-component
system of $^6$Li atoms. For the three lowest hyperfine states of
$^6$Li atoms, the three pair scattering lengths approach a common
large negative value at large magnetic fields and all three have
nearby Feshbach resonances at lower fields that can be used to vary
the scattering lengths (Bartenstein {\it et al.} 2005).  The first experimental
studies of many-body systems of $^6$Li atoms in the three lowest
hyperfine states have been carried out by Ottenstein {\it et al.}\
(2008) and by Huckans {\it et al.}\ (2009).  Their
measurements of the three-body recombination rate revealed a narrow loss
feature and a broad loss feature in a region of low magnetic field.
Theoretical calculations of the three-body recombination rate supported
the interpretation that the narrow loss feature arises from an Efimov
trimer crossing the three-atom threshold (Braaten {\it et al.}  2009;
Naidon \& Ueda 2009;
Floerchinger {\it et al.} 2009).  
Very recently, another narrow loss feature was
discovered in a much higher region of the magnetic field by Williams
{\it et al.}\ (2009) and by Jochim and coworkers. Williams {\it et al.}\
used measurements of the three-body recombination rate in this region to
determine the complex three-body parameter that governs Efimov physics in
this system.  This parameter, together with the three scattering
lengths as functions of the magnetic field, determine the universal
predictions for $^6$Li atoms in this region of the magnetic field. 

Another process that provides signatures for the Efimov effect is atom-dimer
relaxation. This process can occur in a mixture of atoms and shallow dimers. 
There are two types of dimers in ultracold gases: First, there are the
shallow dimers with a binding energy given by the universal formula
\begin{equation}
B_2 = \frac{\hbar^2}{m a^2}\,.
\end{equation}
Second, there are (many) deeply bound dimers with a binding energy 
of order $\hbar^2/m\ell^2$. These dimers are not universal and their
properties depend on the details of physics at short-distances.
The atoms and shallow dimers can undergo inelastic collisions into atoms 
and deeply bound dimers. The difference in the binding energies of the 
shallow dimer and deep dimer is released as kinetic energy and the atom 
and deep dimer in the final state recoil from each other. 
The total relaxation rate into deep dimers 
displays log-periodic scaling and is enhanced
whenever the binding energy of the shallow dimer equals 
the binding energy of an Efimov trimer.  This happens 
when the scattering length
is equal to $(\mathcal{S}_0)^n\,a_*$ with $n$ an integer 
(cf.~Fig.~\ref{fig:efimovplot}). In an 
experiment with a mixture of bosonic $^{133}$Cs atoms and dimers, Knoop et
al.\ observed this resonant enhancement in the loss of atoms and dimers
(Knoop {\it et al.} 2008). This loss feature could be explained by an Efimov
trimer crossing the atom-dimer threshold (Helfrich \& Hammer 2009).

The three-component $^6$Li system provides a new aspect to
atom-dimer relaxation. If at least two of the three atom-atom scattering
lengths in the three-component system are positive, atom-dimer relaxation can
also proceed from the shallow dimer with the smaller binding energy to
the one with the larger binding energy. In $^6$Li, there are in
principle three different relaxation processes possible since there are
three types of shallow dimers: $12$, $23$, and $13$. Additionally, each of
these relaxation rates can be decomposed into the {\it partial} 
relaxation rates into shallow and deep dimers, respectively.

\begin{figure}[t]
\centerline{\includegraphics*[width=11cm,angle=0,clip=true]{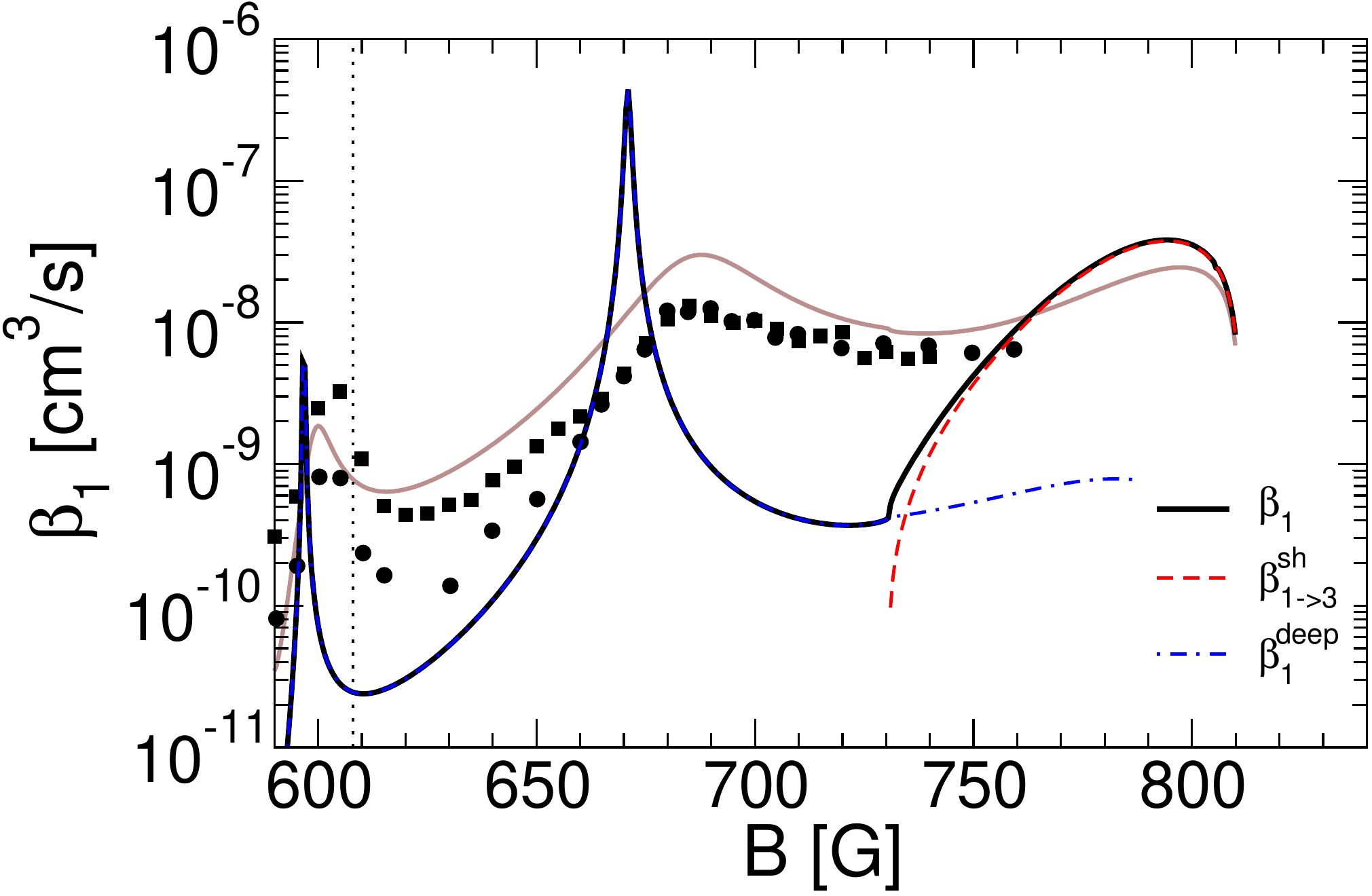}}
\vspace*{0.0cm}
\caption{The relaxation rate constant $\beta_1$ for the 23-dimer and atom 1 as a
  function of the magnetic field $B$.  The squares and circles are data points
  from Lompe 2010 and Nakajima 2010, respectively.  The curves are our results
  for the total rate $\beta_1$ (solid line), the partial rate into atom 3 and
  the 12-dimer $\beta^{\rm sh}_{1\to3}$ (dashed line), and the rate into an
  atom and a deep dimer $\beta^{\rm deep}_1$(dashed-dotted line) using the
  three-body recombination data as input. The light
  solid line gives the total rate $\beta_1$ for the parameters obtained in
  Lompe 2010 while the vertical line marks the boundary of the universal
  region.}
\label{fig:beta1}
\end{figure}
Measurements of the atom-dimer relaxation rate were reported by 
Lompe {\it et al.} (2010a) and Nakajima {\it et al.}
(2010). A theoretical analysis for the relaxation rate of the particular channel
relevant to their experimental setup was performed by Nakajima {\it et al.}
(2010). A complete analysis for all possible partial relaxation rates
was carried out in Hammer {\it et al.} (2010). In Fig. \ref{fig:beta1}, 
we show their results for the relaxation rate constant $\beta_1$ 
for the 23-dimer and atom 1. We compare with the measurements
of Lompe {\it et al.} (2010a) and Nakajima {\it et al.} (2010). 
The figure shows the full relaxation rate as
well the individual contributions from shallow and deep dimers. In the
magnetic field region from $590$~G to $730$~G. While a qualitative agreement 
can be obtained some questions concerning the quantitative description of
the data remain (Hammer {\it et al.} 2010). 
Including non-universal corrections in the 
two- and three-body sector can partially resolve these questions 
(Naidon \& Ueda 2010). 

All experiments discussed so far observed Efimov states indirectly
through their signature in atom loss rates.
Lompe {\it et al.} (2010b) recently reported on the association and 
direct observation of a trimer state consisting of three spin states
of $^6$Li atoms using radio-frequency spectroscopy. Their 
binding energy measurements are consistent with theoretical predictions 
which include non-universal corrections.

\subsection{Nuclear Physics}
\label{sec:nuclear-physics}
The properties of hadrons and nuclei are determined by quantum
chromodynamics (QCD), a non-abelian gauge theory formulated in terms
of quark and gluon degrees of freedom. At low energies, however, the
appropriate degrees of freedom are the hadrons. Efimov physics and
the unitary limit can serve as a starting point for effective
field theories (EFTs) describing hadrons and nuclei at very low
energies.  For convenience, we will now work in natural units where
$\hbar=c=1$.

In nuclear physics, there are a number of EFTs which are useful for a
certain range of systems (Bedaque \& van Kolck 2002; Epelbaum {\it et
  al.} 2009).  At very low
energies, where Efimov physics plays a role, all interactions can be
considered short-range and even the pions can be integrated out. This
so-called {\it pionless} EFT is formulated in an expansion of the
low-momentum scale $\Mlo$ over the high-momentum scale $\Mhi$. It can
be understood as an expansion around the limit of infinite scattering
length or equivalently around threshold bound states. Its breakdown
scale is set by one-pion exchange, $\Mhi\sim M_\pi$, while $\Mlo \sim
1/a \sim k$.  For momenta $k$ of the order of the pion mass $M_\pi$,
pion exchange becomes a long-range interaction and has to be treated
explicitly. This leads to the chiral EFT whose breakdown scale $\Mhi$
is set by the chiral symmetry breaking scale $\Lambda_\chi$. The
pionless theory relies only on the large scattering length and is
independent of the short-distance mechanism generating it.  At leading
order, it is equivalent to the coordinate space formulation
of the three-body problem with zero-range interactions. However, the
ability of the EFT approach to account systematically for finite range
effects makes it particularly interesting to the three-nucleon problem
where the ratio of $\ell/|a|\approx 1/3$. This theory is therefore ideally
suited to unravel universal phenomena driven by the large scattering
length such as limit cycle physics (Braaten \& Hammer 2003;
Mohr {\it et al.} 2005) and the Efimov effect (Efimov 1970) but also to 
describe very low-energy few-nucleon processes to high accuracy.  

In this section, we will
focus on the aspects of nuclear effective field theories related to
Efimov physics.  In the two-nucleon system, the pionless theory
reproduces the well known effective range expansion in the large
scattering length limit. The renormalized S-wave scattering amplitude
to next-to-leading order in a given channel takes the form
(Kaplan {\it et al.} 1998; van Kolck 1999)
\begin{eqnarray}
T (k) &=& \frac{4\pi}{m}
\frac{1}{-1/a-ik} \left[ 1-\frac{r_0 k^2/2}{-1/a-ik}+\ldots \right]\,,
\end{eqnarray}
where $k$ is the relative momentum of the nucleons, $r_0$ denotes the
effective range, and the dots
indicate corrections of order $(\Mlo /\Mhi)^2$ for typical momenta
$k\sim\Mlo$. In the language of the renormalization group, this
corresponds to an expansion around the non-trivial fixed point for
$1/a=0$ (Kaplan {\it et al.} 1998; Birse {\it et al.} 1999). 
The pionless EFT becomes
very useful in the two-nucleon sector when external currents are
considered and has been applied to a variety of electroweak processes.
These calculations are reviewed in detail in Bedaque \& van Kolck 2002.

Since the nucleon-nucleon scattering lengths cannot be changed as in
the case of atoms, the consequences of the limit cycle in the
three-nucleon system cannot be measured directly. The scattering
lengths could be manipulated if the quark masses that are parameters
of the underlying theory of strong interactions (QCD) could be
varied. This is in fact done in numerical simulations of QCD (lattice
QCD) that aim at a calculation of the $NN$ scattering length from
first principles. Simulations with heavier quark masses are
computationally cheaper but their results can be extrapolated to the
physical values using chiral effective field theory. It turns out that
chiral effective field theory indicates the possibility that the $NN$
scattering lengths diverge simultaneously at a quark mass values
slightly larger than the physical ones. Braaten and Hammer conjectured
therefore that a lattice QCD simulation of the three-nucleon system
with quark masses close to this {\it critical} value would therefore
lead to several bound states related to each other by Efimov's scaling
factor of 515 (Braaten \& Hammer 2003). The consequences of this 
infrared limit cycle were studied in the works by Epelbaum {\it et al.} 
(2006) and Hammer {\it et al.} (2007).

Three-body calculations with external currents are still in their
infancy. However, a few exploratory calculations have been carried
out.  Universal properties of the triton charge form factor were
investigated by Platter \& Hammer (2006) and neutron-deuteron
radiative capture was calculated by Sadeghi \& Bayegan (2005), Sadeghi
{\it et al.} (2006), Sadeghi (2007).
Electromagnetic properties of the triton were recently investigated by
Sadeghi \& Bayegan (2010).  This work opens the
possibility to carry out accurate calculations of electroweak
reactions at very low energies for astrophysical processes.

The pionless approach has also been extended to the four-nucleon
sector (Platter {\it et al.} 2005). In order to be able to apply the Yakubovsky
equations, an equivalent effective quantum mechanics formulation was
used. The study of the cutoff dependence of the four-body binding
energies revealed that no four-body parameter is required for
renormalization at leading order.  As a consequence, there are
universal correlations in the four-body sector which are also driven
by the large scattering length.  The best known example is the Tjon
line: a correlation between the triton and alpha particle binding
energies, $B_t$ and $B_\alpha$, respectively. Of course, higher order
corrections break the exact correlation and generate a band.

Another interesting development is the application of the Resonating
Group Model to solve the pionless EFT for three- and four-nucleon
systems (Kirscher {\it et al.} 2010). This method allows for a
straightforward inclusion of Coulomb effects. Kirscher {\it et al.}
extended previous calculations in the four-nucleon system to
next-to-leading order and showed that the Tjon line correlation
persists. Moreover, they calculated the correlation between the singlet
S-wave $^3$He-neutron scattering length and the triton binding
energy. Preliminary results for the halo nucleus $^6$He have also been
reported by Kirscher {\it et al.} (2009).

The pionless theory has also been applied within the no-core shell
model approach. Here the expansion in a truncated harmonic oscillator
basis is used as the ultraviolet regulator of the EFT. The effective
interaction is determined directly in the model space, where an exact
diagonalization in a complete many-body basis is performed. In
(Stetcu {\it et al.} 2007a), the $0^+$ excited state of $^4$He and the
$^6$Li ground state were calculated using the deuteron, triton, alpha
particle ground states as input. The first $(0^+;0)$ excited state in
$^4$He is calculated within 10\% of the experimental value, while the
$^6$Li ground state comes out at about 70\% of the experimental value
in agreement with the 30 \% error expected for the leading order
approximation. These results are promising and should be improved if
range corrections are included.  Finally, the spectrum of trapped
three- and four-body systems was calculated using the same method
(Stetcu {\it et al.} 2007b, 2010; Rotureau {\it et al.} 2010; 
T\"olle {\it et al.} 2010). 
In this case the harmonic potential is physical and not simply used 
as an ultraviolet regulator. 

A special class of nuclear systems exhibiting universal behavior are
{\it halo nuclei} (Zhukov {\it et al.} 1993; Jensen {\it et al.} 2004).  
Halo nuclei consist of
a tightly bound core surrounded by one or more loosely bound valence
nucleons. The valence nucleons are characterized by a very low
separation energy compared to those in the core.  As a consequence,
the radius of the halo nucleus is large compared to the radius of the
core. A trivial example is the deuteron, which can be considered a
two-body halo nucleus. The root mean square radius of the deuteron is
about three times larger than the size of the constituent nucleons.
Halo nuclei with two valence nucleons are particularly interesting
examples of three-body systems.  If none of the two-body subsystems are
bound, they are called {\it Borromean} halo nuclei.  This name is
derived from the heraldic symbol of the Borromeo family of Italy,
which consists of three rings interlocked in such way that if any one
of the rings is removed the other two separate.  The most carefully
studied Borromean halo nuclei are $^6$He and $^{11}$Li, which have two
weakly bound valence neutrons (Zhukov {\it et al.} 1993). In the case of
$^6$He, the core is a $^4$He nucleus, which is also known as the
$\alpha$ particle.  The two-neutron separation energy for $^6$He is
about 1 MeV, small compared to the binding energy of the $\alpha$
particle which is about 28 MeV. The neutron-$\alpha$ ($n\alpha$)
system has no bound states and the $^6$He nucleus is therefore
Borromean. There is, however, a strong P-wave resonance in the
$J=3/2$ channel of $n \alpha$ scattering which is sometimes referred
to as $^5$He.  This resonance is responsible for the binding of
$^6$He. Thus $^6$He can be interpreted as a bound state of an
$\alpha$-particle and two neutrons, both of which are in $P_{3/2}$
configurations.

Because of the separation of scales in halo nuclei, they can be
described by extensions of the pionless EFT. One can assume the core
to be structureless and treat the nucleus as a few-body system of the
core and the valence nucleons. Corrections from the structure of the
core appear in higher orders and can be included in perturbation
theory. Cluster models of halo nuclei then appear as leading order
approximations in this \lq\lq halo EFT''.  A new facet is the
appearance of resonances as in the neutron-alpha system which leads to
a more complicated singularity structure and renormalization compared
to the few-nucleon system discussed above (Bertulani 2002).

The first application of effective field theory methods to halo nuclei 
was carried out by Bertulani {\it et al.} (2002) and 
Bedaque {\it et al.} (2003), where the 
$n\alpha$ system (``$^5$He'') was considered. They found that 
for resonant P-wave interactions both the scattering length and effective 
range have to be resummed at leading order. In this case, two 
coupling constants in the effective Lagrangian are fine tuned to
unnatural values. At threshold, however, only one fine tuning is
required and the EFT becomes perturbative. Because the $n\alpha$ 
interaction is resonant in the P-wave 
and not in the S-wave, the binding mechanism of $^6$He is not the Efimov
effect. However, this nucleus can serve as a laboratory for studying
the interplay of resonance structures in higher partial waves. 

Three-body halo nuclei composed of a core and two valence neutrons are
of particular interest due to the possibility of these systems to
display the Efimov effect (Efimov 1970). Since the scattering
length can not be varied in halo nuclei, one looks for Efimov
scaling between different states of the same nucleus. Such analyses
assume that the halo ground state is an Efimov state.\footnote{We note
  that it is also possible that only the excited state is an Efimov
  state while the ground state is more compact. This scenario can not
  be ruled out but is also less predictive.} They have previously been
carried out in cluster models and the renormalized zero-range model
(See, e.g. Federov {\it et al.} 1994; Amorim {\it et al.} 1997; 
Mazumdar {\it et al.} 2000).
A comprehensive study of S-wave halo nuclei in halo EFT was
carried out by Canham \& Hammer (2008).  This work provided binding
energy and structure calculations for various halo nuclei including
error estimates. Confirming earlier results by Fedorov et
al. (1994) and Amorim {\it et al.}~(Amorim {\it et al.} 1997),
$^{20}$C was found to be the only candidate nucleus for an excited
Efimov state assuming the ground state is also an Efimov state.  This
nucleus consists of a $^{18}$C core with spin and parity quantum
numbers $J^P=0^+$ and two valence neutrons. The nucleus $^{19}$C is
expected to have a $\frac{1}{2}^+$ state near threshold, implying a
shallow neutron-core bound state and therefore a large neutron-core
scattering length. The value of the $^{19}$C energy, however, is not
known well enough to make a definite statement about the appearance of
an excited state in $^{20}$C. An excited state with a binding energy
of about 65~keV is marginally consistent with the current experimental
information.

\begin{figure}[tb]
\centerline{\includegraphics*[width=10cm,angle=0]{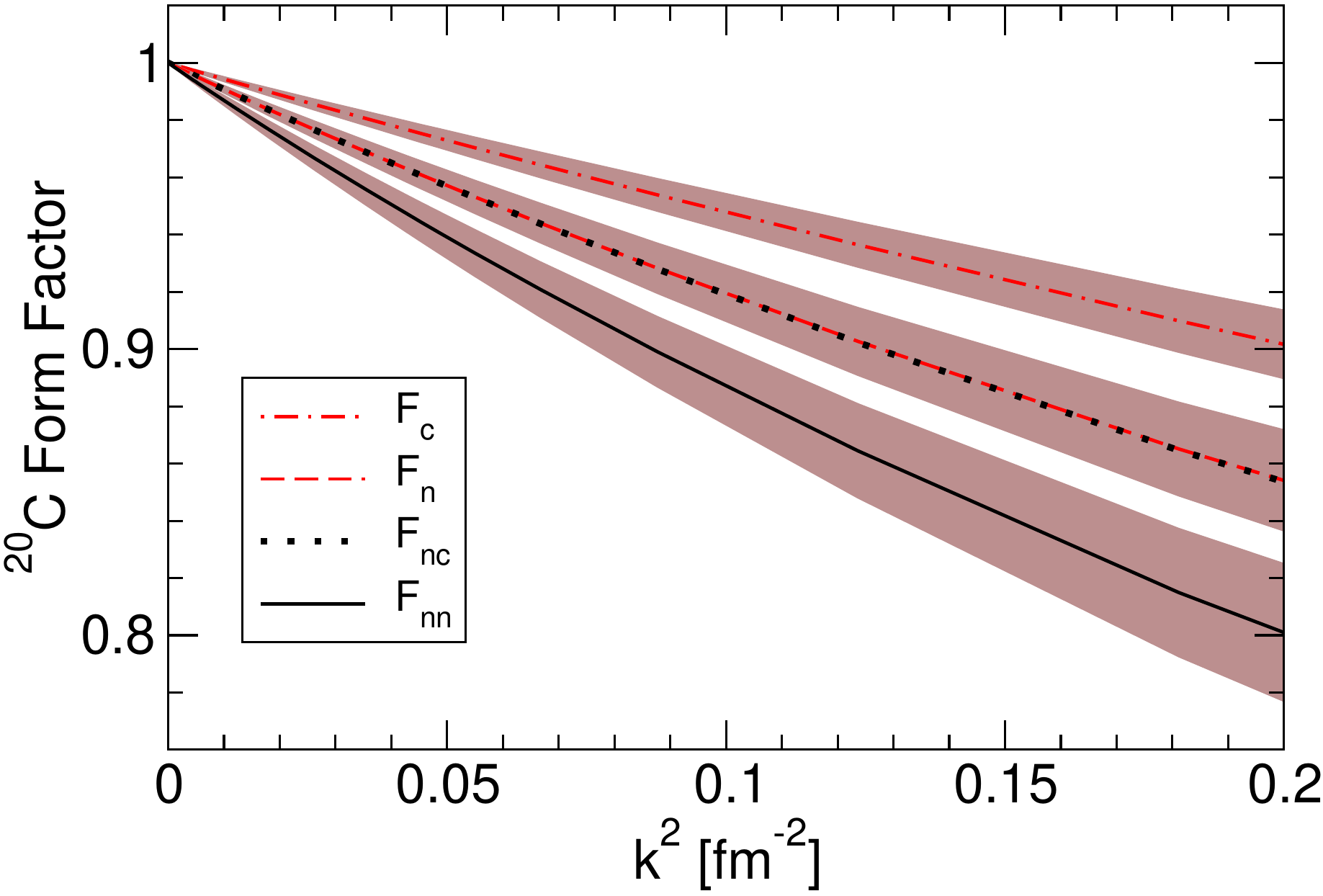}}
\caption{The one- and two-body matter density form factors 
${\cal F}_{c}$, ${\cal F}_{n}$, ${\cal F}_{nc}$, and 
${\cal F}_{nn}$ with leading order error bands for the ground 
state of $^{20}$C as a function of the momentum transfer $k^2$.}
\label{fig:spec.halo}
\end{figure}

The matter form factors and radii of halo nuclei can also be
calculated in the halo EFT (Yamashita {\it et al.} 2004; 
Canham \& Hammer 2008).  As
an example, we show the various one- and two-body matter density form
factors ${\cal F}_{c}$, ${\cal F}_{n}$, ${\cal F}_{nc}$, 
and ${\cal F}_{nn}$
with leading order error bands for the ground state of $^{20}$C as 
a function of the momentum transfer $k^2$ from Canham \& Hammer (2008) 
in Fig.~\ref{fig:spec.halo}.
The effective theory description breaks down for 
momentum transfers of the order of the pion-mass 
squared ($k^2\approx 0.5$ fm$^{-2}$). 
The range corrections to these results were estimated in
(Canham \& Hammer 2010) and found to be generally small.

Scattering observables offer a complementary window on Efimov physics
in halo nuclei and some recent model studies have focused on this
issue.  In the works by Yamashita {\it et al.} (2007) and Mazumdar
{\it et al.} (2006) the trajectory of the
possible $^{20}$C excited state was extended into the scattering
region in order to find a resonance in $n$-$^{19}$C scattering. A
detailed study of $n$-$^{19}$C scattering near an Efimov state was
carried out by Yamashita {\it et al.} (2008).

\section{Outlook and Conclusion}
In this work we have discussed the Efimov effect in the context of the
renormalization group. The discrete scaling
invariance particular to the Efimov effect 
can be understood as the consequence of an RG limit cycle. The
log-periodic scattering length dependence of observables is therefore
a prominent signature of the limit cycle in experimental
observables. We have also reviewed the steadily growing number of
observations of Efimov physics in ultracold atomic gases. In
particular, the ability to precisely tune the scattering length 
using Feshbach resonances has
made it possible to observe the log-periodic scaling in three-body
loss processes. 

The situation in nuclear physics is very different since the 
value of the scattering length
is fixed. Only a consistent description of different three-body observables
within the same universal EFT approach can therefore be regarded as evidence
for Efimov physics. A number of such calculations has been 
performed. In particular, the properties of the triton have been thoroughly
analyzed in the pionless EFT. These calculations include various
electromagnetic, scattering and bound state properties. It is furthermore
important to note that few-nucleon physics is not only
interesting because of the universal aspects but also has important practical
applications.  Low-energy energy nuclear reactions are hard to
measure but their knowledge is important for the description of
astrophysical processes such as big bang nucleosynthesis. This aspect requires
a consistent inclusion of higher order corrections, e.g. from finite
range effects, and external currents.

Future lattice QCD simulations of few-nucleon systems might be
able to provide more evidence for the possible existence of an infrared limit
cycle in QCD (Kreuzer \& Hammer 2010).  In this approach, 
the observables are extracted from correlation 
functions calculated by evaluating the QCD path integral on a Euclidean
space time lattice using Monte Carlo methods.
With high statistics Lattice QCD simulations of
three-baryon systems within reach (Beane {\it et al.} 2010), 
the calculation of the
structure and reactions of light nuclei directly from QCD appears feasible in
the intermediate future.

Halo nuclei are an additional class of systems that might provide examples 
of Efimov states.  They consist of a tightly bound core surrounded 
by one or more weakly bound valence nucleons and can be considered
effective few-body systems. 
The properties of a number of three-body halo nuclei have been
analyzed. In particular, bound state properties such as matter radii and form
factors have been calculated. In the two-body sector various scattering
and breakup processes have been considered.
In the future, the effective field theory approach might 
also be helpful in analyzing reactions involving three-body halo 
nuclei since it reduces the
number of degrees of freedom included in a controlled manner and 
higher order corrections can be included systematically.

Since the discovery of the Efimov effect approximately forty years
ago, it has
fascinated an ever growing number of physicists. It is not only a remarkable
example of discrete scale invariance on the quantum level but it also relates
physical systems across different branches of physics including,
atomic, nuclear and particle physics. It serves as paradigm for a 
universal binding mechanism in weakly-bound few-body systems with 
binding energies from neV to MeV.  Furthermore, it provides a starting
point for an effective field theory that can facilitate the calculation of
low-energy observables to high accuracy.

\section*{Acknowledgements}
This research was supported by the Department of Energy under grant
number DE-FG02-00ER41132, by the Deutsche Forschungsgemeinschaft
through SFB/TR16, and by the Bundesministerium f\"ur Bildung und
Forschung under contract no. 06BN9006.

\end{document}